\documentclass[8pt]{article}
\usepackage{graphicx}
\usepackage{pgfplots}
\usepackage{newlfont}
\usepackage[colorlinks,linkcolor=blue,citecolor=blue,anchorcolor=blue,bookmarksopen,pdfpagetransition={Wipe}]{hyperref}
\usepackage{amsmath,amsthm,amssymb}
\usepackage{subfigure}
\usepackage{amsmath}
\usepackage{booktabs}
\usepackage{fourier}
\usepackage{array}
\usepackage{makecell}
\usepackage{algorithm}
\usepackage{algpseudocode}
\usepackage{xcolor}

\newtheorem{thm}{Theorem}[section]

\newtheorem{rem}[thm]{Remark}
\numberwithin{equation}{section}

\DeclareMathOperator{\ultang}{ULTS1D}
\DeclareMathOperator{\fcs}{FCS1D}
\DeclareMathOperator{\scs}{SCS1D}
\DeclareMathOperator{\std}{S3D}
\DeclareMathOperator{\invfcs}{InvFCS1D}
\DeclareMathOperator{\invscs}{InvSCS1D}
\DeclareMathOperator{\invstd}{InvS3D}
\DeclareMathOperator{\fliplr}{fliplr}
\DeclareMathOperator{\flipud}{flipud}
\DeclareMathOperator{\Sz}{size}

\DeclareMathOperator{\reshape}{reshape}
\DeclareMathOperator{\circshift}{circshift}
\DeclareMathOperator{\key}{key}
\DeclareMathOperator{\rand}{rand}
\DeclareMathOperator{\mean}{mean}
\DeclareMathOperator{\sort}{sort}
\DeclareMathOperator{\diag}{diag}
\begin{document}
\title{\textbf{Four-dimensional hybrid chaos system and its application in creating a secure image transfer environment by cellular automata}}
\author{
R. Parvaz $^a$\footnote{ rparvaz@uma.ac.ir,~reza.parvaz@yahoo.com}\
, Y. Khedmati$^a$\footnote{Corresponding author:khedmati.y@uma.ac.ir,~khedmatiy.y@gmail.com}\
, Y. Behroo $^a$\footnote{yousef.behroo@uma.ac.ir}
}
\date{}
\maketitle
\begin{center}
$^a$Department of Mathematics, University of Mohaghegh Ardabili,
56199-11367 Ardabil, Iran.\\
\end{center}
\begin{abstract}
\noindent
One of the most important and practical researches which has been
considered by researchers is creating secure environments for information exchanges.
Due to their structures, chaos systems are efficient tools in the are of data transferring.
In this research, using a mathematical structure such as composing and transferring, we improve classical chaotic systems by creating a four-dimensional system. Then we introduce a new encryption algorithm based on the chaos and cellular automata.
The security of the proposed environment which is evaluated using different types of security tests shows the efficiency of the proposed algorithm.
\end{abstract}
\vskip.3cm

\section{Introduction}
With the rapid development of social networks, a huge amount of various information in the form of text, image, audio and video
is exchanged through these networks in any small period of time.
Because of their visual features, digital images
have been received more attention among other formats, and so
it is important to present a new method or improve existing methods in order to transfer images safely.
In addition to steganography, cryptography is a method of secure transfer of information in which the goal is to scramble pixels of data  properly
so that it is not detectable \cite{ii1}. So far, various cryptographic methods have been proposed based on cellular automata, chaotic mapping, and so on for example see \cite{ii2,ii3,ii4}. Chaos systems are nonlinear phenomena with random-like behaviors. These maps are very important in information security transforms with due regard to their special features, which we will discuss in detail in future sections.
Given that researchers have shown that algorithms that use existing chaotic maps are likely to be attacked \cite{i51}, defining appropriate maps is a priority for cryptographic methods. The dimension of a chaos maps are determined based on the number of input and output variables. In general, it can be said that high-dimensional chaotic maps, although having high computational cost, perform better than one-dimensional maps due to their complex dynamic structure.
\par Cellular automata is one of the image encryption tools that has been used in this article due to its fast, easy and high speed process.
In \cite{i32}, Ray et al use chaos and 3D automation to encode digital images.
First, the replacement operation is performed by a chaotic system, and then the diffusion operation is performed using a three-dimensional cellular automata. This method has shown good resistance to most known cryptographic attacks.
Wang's method is an image encryption system based on chaos theory and cellular automation.
This method uses a logistic map and reversible cellular automata. Pixel values divided in four-bit units, then in the permutation step of logistic map and the distribution step, cellular automation is also used. This method of cryptography, which is one of the symmetric methods, has good security and shows good resistance against differential attacks \cite{i33}.

\section{4D Hybrid Chaos Systems}
In the first step of this section, we introduce a novel type of four-dimensional chaotic system by improving its overall structure, and then in the second subsection, we evaluate the proposed chaos behavior using the various tests such as
lyapunov exponent and cobweb diagram.
\subsection{Structure of the Proposed Hybrid System}
\begin{figure}[h!]
\centering
\subfigure{
\includegraphics*[width=1\textwidth]{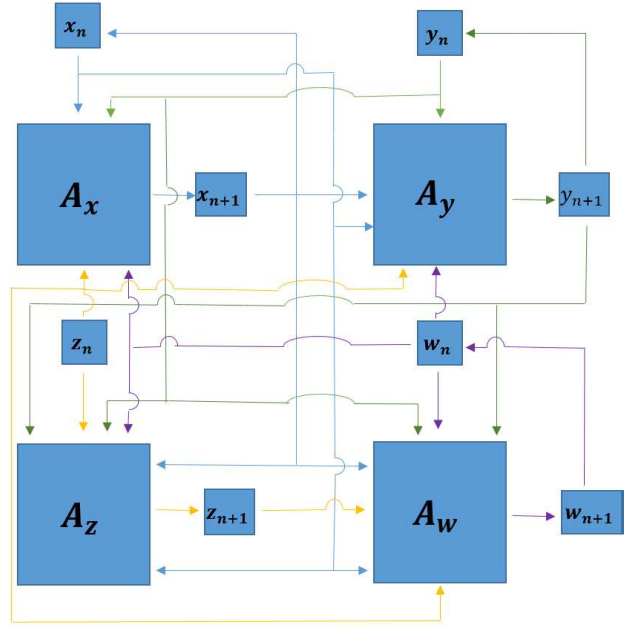}}\\
\vspace{-1.8cm}
\emph{\caption{
The structure of the proposed chaotic system.}\label{fs1}}
\end{figure}

\begin{figure}[h!]
\centering
\subfigure{
\includegraphics*[width=1\textwidth]{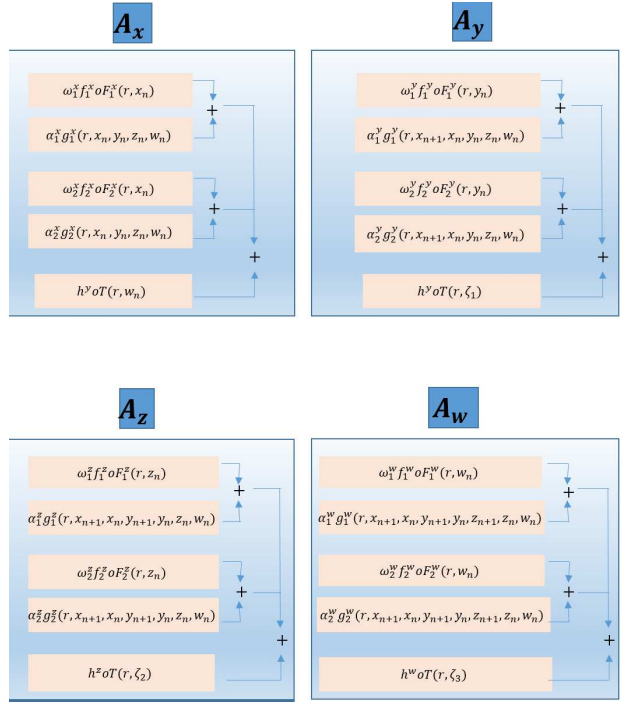}}\\
\vspace{-1.8cm}
\emph{\caption{
The structure of the combination parts in the proposed system. }\label{f2}}
\end{figure}

In this subsection, more details about the proposed new hybrid chaos systems based on Tent, Sin and Logistic maps are given. The general structure of the proposed chaos system has been given in Fig. \ref{fs1}. The combination parts in the proposed system based on Tent, Sin and Logistic maps are shown in Fig. \ref{f2}. The mathematics formulae for each of the parts can be written as follows.

\newpage

\noindent-First combination box:
\begin{align}
x_{i+1}:=
\left\{%
\begin{array}{ll}
\alpha^x_1f^x_1 \circ F^x_1(r,x_i)+g^x_1(r,x_i,y_i,z_i,w_i)\\
~~~~~~~~~~~~+h_1^x(\frac{(\beta^x_1-r)z_{i}}{2})~mod~1,~~when~w_{i}<0.5,\\
\\
\alpha^x_2f^x_2 \circ F^x_2(r,x_i)+g^x_2(r,x_i,y_i,z_i,w_i)\\
~~~~~~~~~~~~+h_2^x(\frac{(\beta^x_2-r)(1-z_{i})}{2})~mod~1,~~when~w_{i}\geq0.5.\\
\end{array}%
\right.
\end{align}
\noindent-Second combination box:
\begin{align}
y_{i+1}:=
\left\{%
\begin{array}{ll}
\alpha^y_1f^y_1 \circ F^y_1(r,y_i)+g^y_1(r,x_{i+1},x_i,y_i,z_i,w_i)\\
~~~~~~~~~~~~+h_1^y(\frac{(\beta^y_1-r)\zeta_1}{2})~mod~1,~~when~\zeta_1<0.5,\\
\\
\alpha^y_2f^y_2 \circ F^y_2(r,y_i)+g^y_2(r,x_{i+1},x_i,y_i,z_i,w_i)\\
~~~~~~~~~~~~+h_2^y(\frac{(\beta^y_2-r)(1-\zeta_1)}{2})~mod~1,~~when~\zeta_1\geq0.5.\\
\end{array}%
\right.
\end{align}
\noindent-Third combination box:
\begin{align}
z_{i+1}:=
\left\{%
\begin{array}{ll}
\alpha^z_1f^z_1 \circ F^z_1(r,w_i)+g^z_1(r,x_{i+1},x_i,y_{i+1},y_i,z_i,w_i)\\
~~~~~~~~~~~~+h_1^z(\frac{(\beta^z_1-r)\zeta_2}{2})~mod~1,~~when~\zeta_2<0.5,\\
\\
\alpha^z_2f^z_2 \circ F^z_2(r,w_i)+g^z_2(r,x_{i+1},x_i,y_{i+1},y_i,z_i,w_i)\\
~~~~~~~~~~~~+h_2^z(\frac{(\beta^z_2-r)(1-\zeta_2)}{2})~mod~1,~~when~\zeta_2\geq0.5.\\
\end{array}%
\right.
\end{align}
\noindent-Fourth combination box:
\begin{align}
w_{i+1}:=
\left\{%
\begin{array}{ll}
\alpha^w_1f^w_1 \circ F^w_1(r,z_i)+g^w_1(r,x_{i+1},x_i,y_{i+1},y_i,z_{i+1},z_i,w_i)\\
~~~~~~~~~~~~+h_1^w(\frac{(\beta^w_1-r)\zeta_3}{2})~mod~1,~~when~\zeta_3<0.5,\\
\\
\alpha^w_2f^w_2 \circ F^w_2(r,z_i)+g^w_2(r,x_{i+1},x_i,y_{i+1},y_i,z_{i+1},z_i,w_i)\\
~~~~~~~~~~~~+h_2^w(\frac{(\beta^w_2-r)(1-\zeta_3)}{2})~mod~1,~~when~\zeta_3\geq0.5,\\
\end{array}%
\right.
\end{align}
\noindent where
$a_1:=\{r,x_i,y_i,z_i,w_i\},$
$ a_2:=\{r,x_i,x_{i+1},y_i,z_i,w_i\}$,
$a_3:=\{r,x_i,x_{i+1},y_i,y_{i+1},z_i,w_i \}$,
$a_4:=\{r,x_i,x_{i+1},y_i,y_{i+1},z_i,z_{i+1},w_i\}$, and
$\xi_\tau=\tau_i$  or
$\tau_{i+1},$ for
$\tau=x,y,z.$
$F_\varsigma^\tau,$ for
$\tau=x,y,z,w ,\varsigma=1,2$, can be considered as Sin or Logistic maps. $\alpha_\varsigma^\tau,\beta_\varsigma^\tau,$ for $\tau=x,y,z,w ,\varsigma=1,2,$ are arbitrary number in R, and $g_\varsigma^\tau,h_\varsigma^\tau$ are considered as arbitrary sufficiently smooth functions.  In the proposed system, the best feature of the different chaos maps as Tent, Sin and Logistic maps have been improved by using Composition and transfer operator. In the following, the basic properties of the hybrid system have been studied.
In order to study hybrid system, the following cases have been considered.\\

\textbf{Case i:}
$\{\alpha_1^x,\alpha_2^x,\alpha_1^y,\alpha_2^y,\alpha_1^z,\alpha_2^z,\alpha_1^w,\alpha_2^w \}$
$=\{1,16,10,20,10,20,10,20\}$,
$\{\beta_1^x,\beta_2^x,\beta_1^y,\beta_2^y,$
$\beta_1^z,\beta_2^z,\beta_1^w,\beta_2^w \}$
$=\{6,2,50,30,50,30,50,30\}$,
$\{\xi_x,\xi_y,\xi_z \}$
$=\{x_i,y_i,z_i\}$,
$f_1^x(p)=\cosh(p)$,
$f_2^x(p)=\cot(p)$,
$f_1^y(p)=f_1^z(p)=f_1^w(p)=p$,
$f_2^y(p)=f_2^w(p)=\sin(\pi p)$,
$f_2^z (p)=\exp(\pi p),$
$g_1^x(a_1)=15\tanh(rx_i+z_i)+\sin((w_i)+12 \cos(rx_i)$,
$g_2^x (a_1)=-7ry_i+\exp(1+2w_i )+z_i+7 \log(\pi rx_i )$,
$g_1^y(a_2)=2\tan(rx_i+y_i+2z_i+w_i),$
$ g_2^y(a_2)= z_i+w_i+14 \exp(20rx_i )$,
$g_1^z(a_3)=2\tan(rx_i+y_i)+w_i+z_i,$
$ g_2^z(a_3 )=14 \exp(20rx_i+w_i)+\sin(z_i)$,
$g_1^w(a_4)=2 \tan(rx_i+y_i+z_i )$
$+w_i,g_2^w(a_4)=14\exp(20rx_i+w_i)$
$+z_i,h_1^x(p)=\sin(2p)$
,$h_2^x(p)=4p,h_2^y(p)=\cot(p)$,
$h_1^y(p)=h_1^z(p)=h_1^w(p)=\exp(2p)$,
$h_2^z(p)$
$=h_2^w(p)=\cot(4p)$.\\

\textbf{Case ii: }
$\{\alpha_1^x,\alpha_2^x,\alpha_1^y,\alpha_2^y,\alpha_1^z,\alpha_2^z,\alpha_1^w,\alpha_2^w \}$
$=\{7,12,14,14,3,15,15,10\}$,
$\{\beta_1^x,\beta_2^x,\beta_1^y, \beta_2^y,$
$\beta_1^z,\beta_2^z,\beta_1^w,\beta_2^w\}=$
$\{69,28,68,36,33,2,5,7\}$,
$\{\xi_x,\xi_y,\xi_z\}=$
$\{x_{i+1},y_{i+1},z_{i+1}\}$,
$f_1^x(p)=\cos(p)$,
$f_2^w(p)=\cos(\sin(\pi p))$,
$f_2^x(p)=f_2^y(p)=p$,
$f_1^z(p)=\sin(p)$,
$f_1^w(p)=\sin(\pi p)$,
$f_2^z(p)=\exp(\pi p),$
$g_1^x(a_1)=15\tan(rw_i+x_i+2z_i)+\sin(w_i)+12\cos(rx_i)$,
$g_2^x(a_1)=7\sin(ry_i+w_i)-7ry_i+x_i+2w_i+z_i-1$,
$g_1^y(a_2)=14\cos(20rx_i+x_{i+1})$,
$g_1^z(a_3)=\tan(x_{i+1}+y_{i+1})+w_i$
$z_i+2(rx_i+y_i)$,$g_2^z(a_3)$
$=14(rx_i+w_i)+y_{i+1}+\sin(z_i)$,
$g_1^w(a_4)=14\cos(20rx_i+x_{i+1})$
$+\log(x_i+w_i)$,
$g_2^w(a_4)=14\sinh(x_i+rx_{i+1}+w_i)$
$+z_i+\sin(z_{i+1}+y_{i+1})$,
$h_1^x(p)=\cos(20p),h_1^x(p)=\sin(2p),$
$h_1^y(p)=\exp(2p)$,
$h_2^x(p)=h_2^y(p)=4p$,
$h_1^z(p)=\cosh(2p),$
$h_1^w(p)=\exp(4p)$,
$h_2^z(p)=h_2^w(p)=\coth(p)$.\\

\noindent Also, in the case i, $\{F_1^x,F_1^z,F_2^z,F_2^w\}$ are considered as Logistic map, and $\{F_2^x,F_1^y,F_2^y,$
$F_2^z,F_1^w\}$ are considered as Sin map. For case ii, $\{F_1^x,F_2^y,F_1^z,$
$F_2^z,F_1^w,F_2^w\}$ and
$\{F_2^x,F_1^y\}$ are considered as Logistic and Sin maps, respectively.

\subsection{Chaotic Behavior Analysis}
\begin{figure}[h!]
\centering
\subfigure{
\includegraphics*[width=1\textwidth]{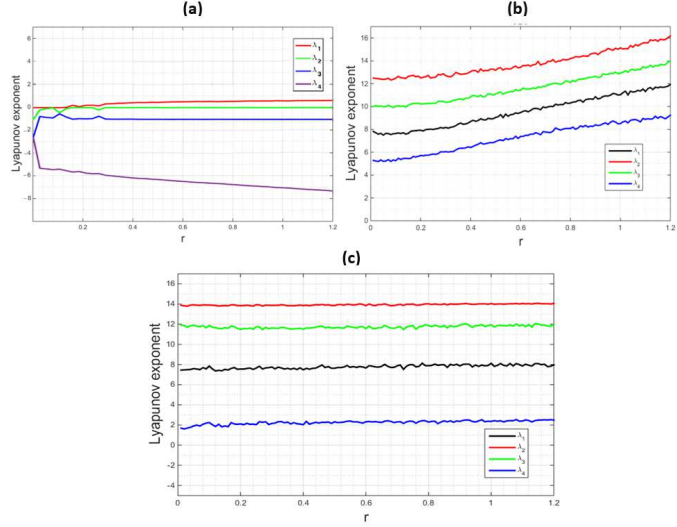}}\\
\vspace{-1.8cm}
\emph{\caption{
Lyapunov exponent values for: (a) Chaotic system in [6], (a) Case i, (b) Case ii.  }\label{f3}}
\end{figure}

In this subsection some important tests for the proposed chaos system are discussed. One of the important value in the study of the behavior of the chaos system is Lyapunov exponent or Lyapunov characteristic exponent. An $n$-dimensional chaos systems in general have n values for Lyapunov exponent. There are many methods for calculating this value \cite{bbn1,bbn2,bbn3}. The method based on QR algorithm has been used for obtained Lyaponov exponent in Fig. \ref{f3} for the case i. More details about this method can be found in \cite{bbn3}. The positive or negative values of the resulting values are related to the structure of a chaos system. This relation had been studied in many papers. In \cite{bbn4}, the relation has been given as follows ``In an n-dimensional dynamical system we have n Lyapunov exponents. Each $\lambda_k$ represents the divergence of k-volume. The sign of the Lyapunov exponents indicates the behavior of nearby trajectories. A negative exponent indicates that neighboring trajectories converge to the same trajectory A positive exponent indicates that neighboring trajectories diverge \cite{bbn4}''. Also the following theorem has been given in \cite{bbn5} for this value.\\

\begin{thm}
 If at least one of the average Lyapunov exponents is positive, then the system is chaotic; if the average Lyapunov exponent is negative, then the orbit is periodic and when the average Lyapunov
exponent is zero, a bifurcation occurs.
\end{thm}

\begin{figure}[h!]
\centering
\subfigure{
\includegraphics*[width=1\textwidth]{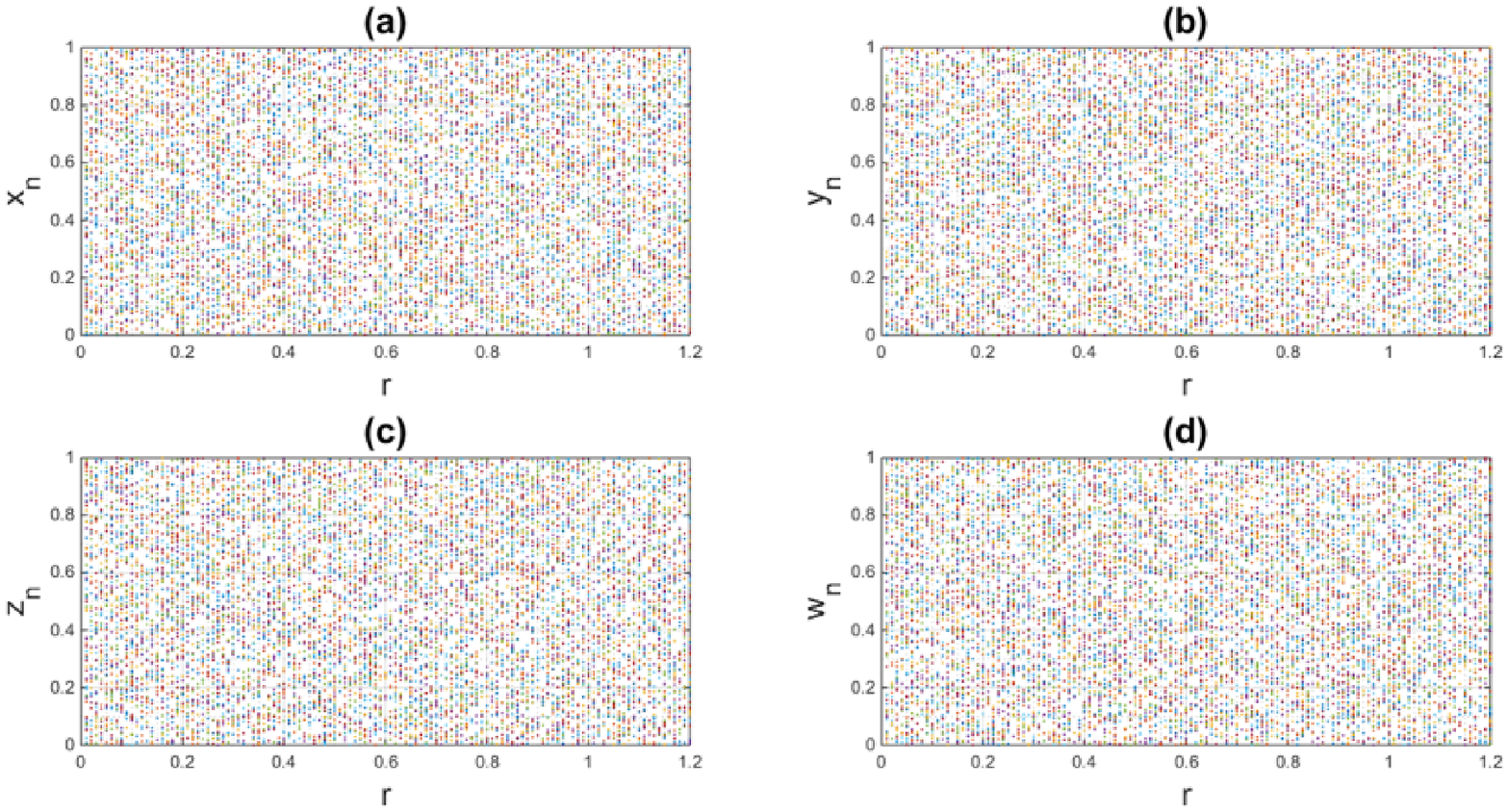}}\\
\vspace{-1.8cm}
\emph{\caption{
Bifurcation diagram results for the case i.}\label{f4}}
\end{figure}

\begin{figure}[h!]
\centering
\subfigure{
\includegraphics*[width=1\textwidth]{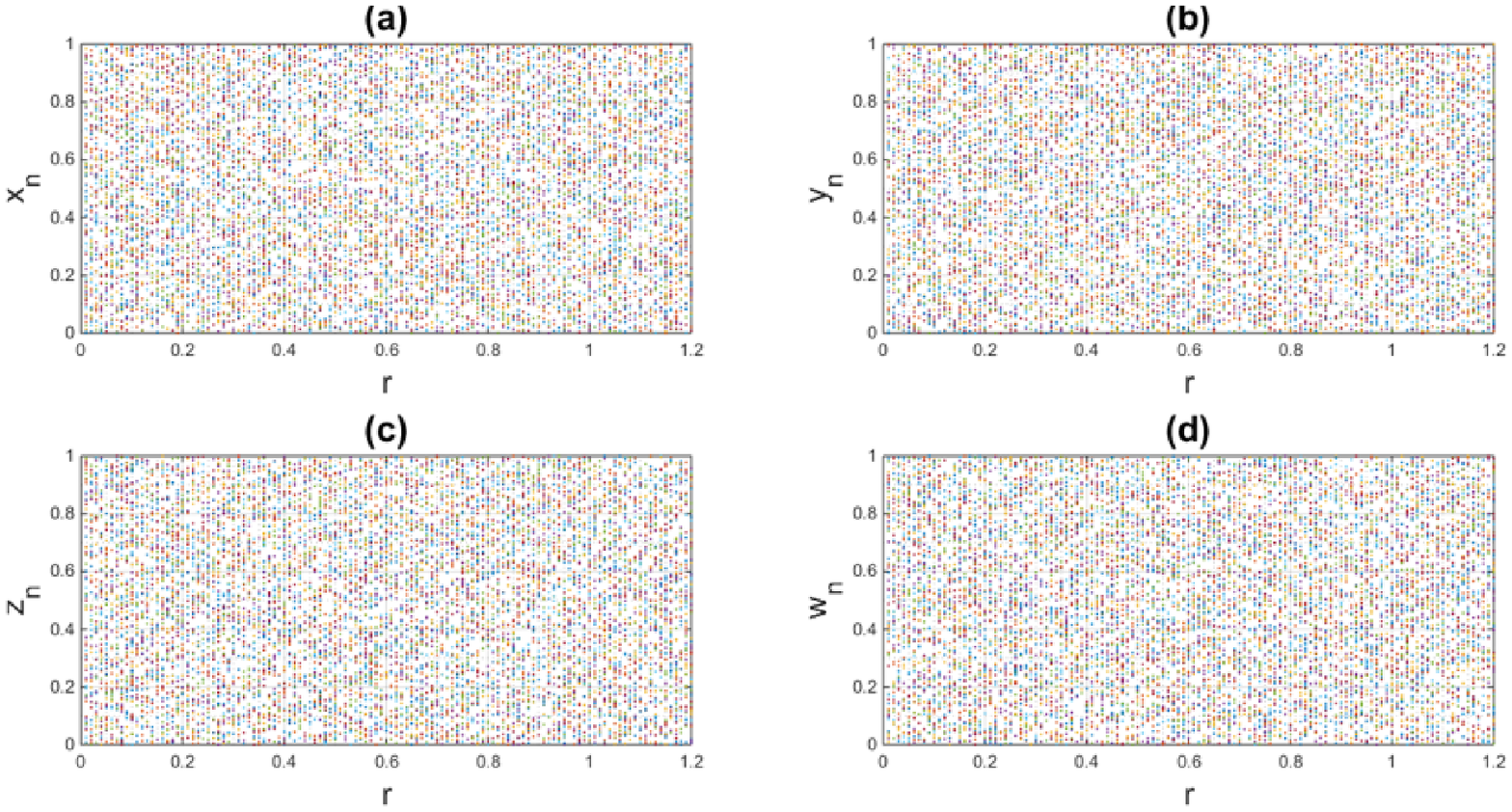}}\\
\vspace{-1.8cm}
\emph{\caption{
Bifurcation diagram results for the case ii.   }\label{f5}}
\end{figure}

\begin{figure}[h!]
\centering
\subfigure{
\includegraphics*[width=1\textwidth]{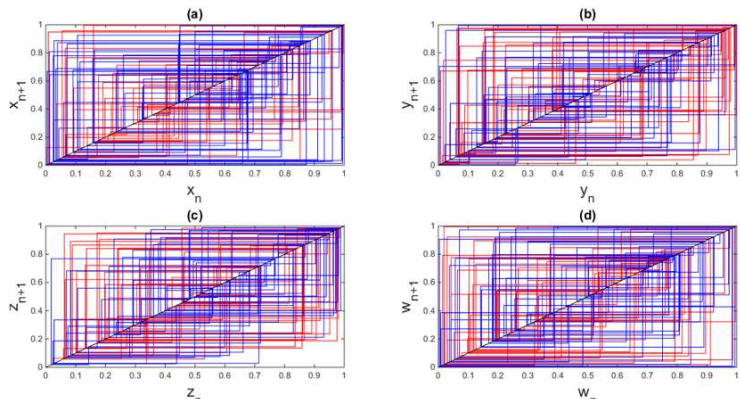}}\\
\vspace{-1.8cm}
\emph{\caption{
Cobweb plots for the case i.  }\label{f6}}
\end{figure}

The results in Fig. \ref{f3} show that all four value of Lyapunov exponent in the proposed systems are positive. By using above studies, we can say that the proposed system in the all neighboring trajectories
diverge. In order to compare proposed system, the Lyapunov exponent has been compared with 4D Chaotic Laser System \cite{bbn6} in Fig. \ref{f3}. It is observed that the proposed system has better chaos behavior than Chaotic Laser System.
Another tool for study chaotic behavior is bifurcation analysis. In the Fig.s \ref{f4}-\ref{f5}, the results for the bifurcation analysis of the case i and ii of the proposed system have been shown.  The chaotic attractors can be studied by this figure. The attractor for $r \in (0 ,1.2]$ is given in the vertical line at the chosen $r$. Also, the cobweb plot (or Verhulst diagram) for case i have been given in the Fig. \ref{f6}. By using this results, it is observed that for the given values, the resulting sequences of the proposed system has chaotic behavior.

\begin{figure}[h!]
\centering
\subfigure{
\includegraphics*[width=1\textwidth]{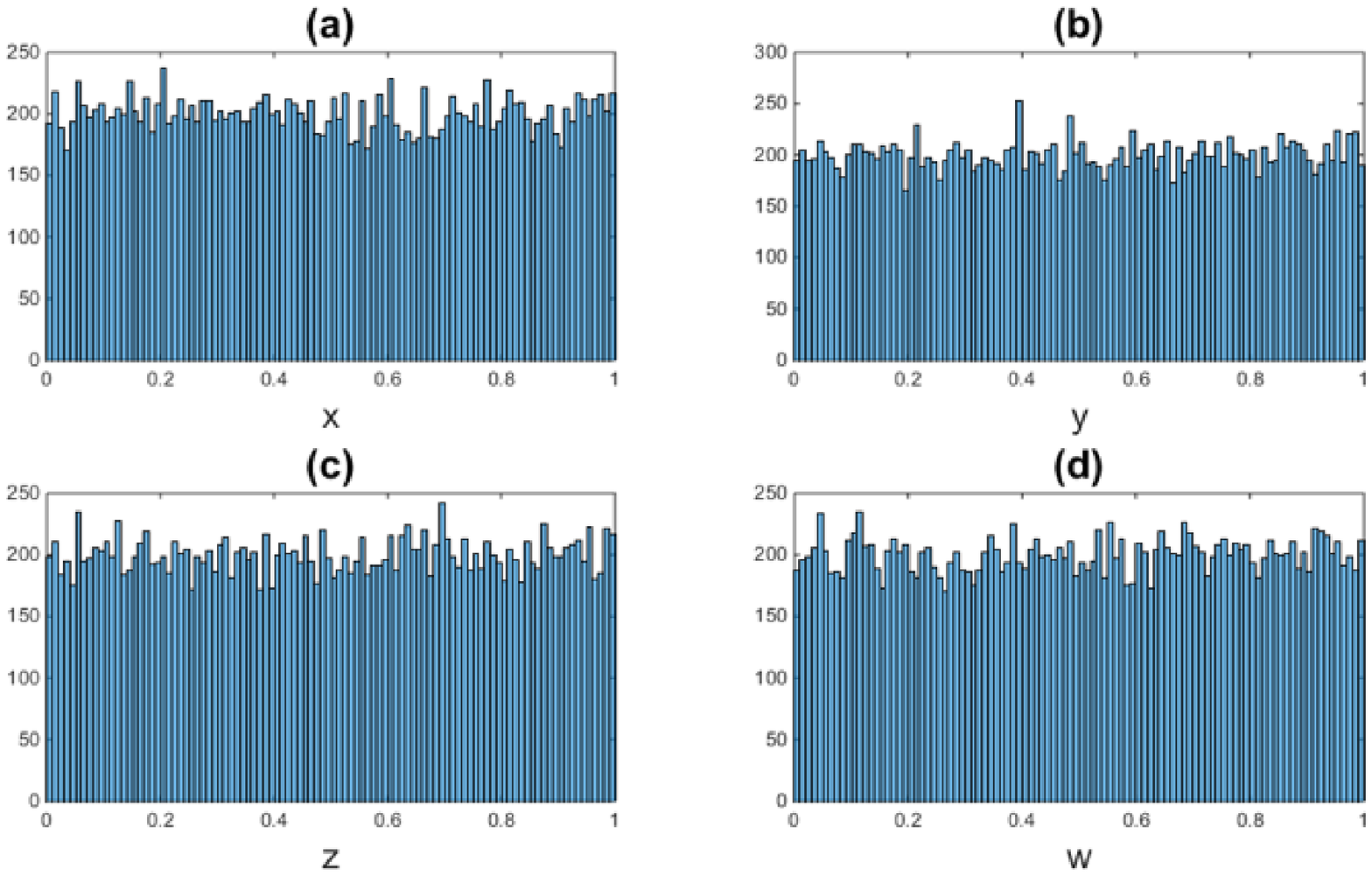}}\\
\vspace{-1.8cm}
\emph{\caption{
Histogram plots of the case i for r=0.5.}\label{f7}}
\end{figure}

\begin{figure}[h!]
\centering
\subfigure{
\includegraphics*[width=1\textwidth]{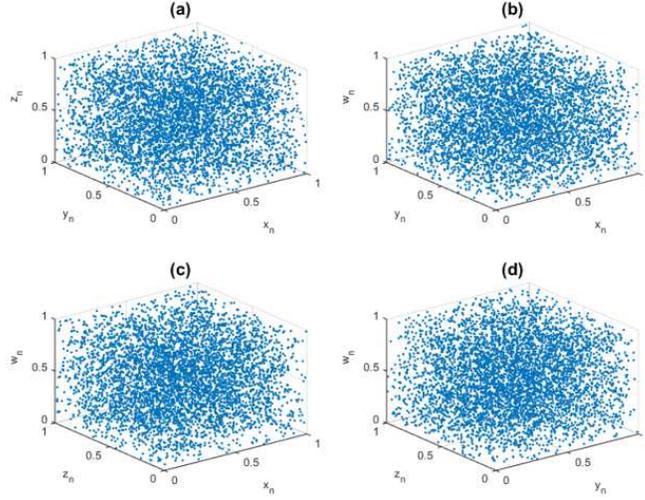}}\\
\vspace{-1.8cm}
\emph{\caption{
Distribution patterns of the case ii for $r=0.4$.  }\label{f8}}
\end{figure}

Distribution is another important factor in evaluating chaotic system. One of the reasons for the weakness of
chaos systems against the statistical attack is nonuniform distribution. The histogram plots of the proposed system for the case i are given in the Fig. \ref{f7}. Also, the distribution patterns of the case ii are shown in Fig. \ref{f8}. By using these results, it can be seen that the generated sequence of the proposed maps have flat distributions.

\begin{rem}
In the continuation of this paper outputs of the proposed four-dimensional chaos system with initial values
$\gamma=(x_0,y_0,z_0,w_0)$ and $r$ is shown by using the following notation
\begin{align*}
^{\gamma}\Psi^n_r := \left[ {\begin{array}{*{20}{c}}
   {\begin{array}{*{20}{c}}
   {x_1}  \\
   {y_1}  \\
   {z_1}  \\
   {w_1}  \\
   \end{array}} & {\begin{array}{*{20}{c}}
    {x_2}  \\
   {y_2}  \\
   {z_2}  \\
   {w_2}  \\
\end{array}} & {\begin{array}{*{20}{c}}
   {\cdots}  \\
   {\cdots}  \\
   {\cdots}  \\
   {\cdots}  \\
\end{array}} & {\begin{array}{*{20}{c}}
    {x_n}  \\
   {y_n}  \\
   {z_n}  \\
   {w_n}  \\
\end{array}}  \\
\end{array}} \right].
\end{align*}
Also ${}_{\tau}^{\gamma}\Psi^n_r $ denotes $^{\gamma}\Psi^n_r $, which the decimal part of the numbers is cut from the $\tau$-th decimal number to the next digits.

\end{rem}

\section{One and three dimensional shift Functions}
In this section, we introduce two types of functions named $\ultang$ and $\std$ to intermix pixels of images.
The former is upper and lower triangular shift function which is suitable for grayscale images and the latter is suitable for RGB images.
Note that applying these functions on images does not change the original image size and only the pixel locations are moved.
In the continuation of this section, $n_1, n_2, n_3, n_4$ are integers and $n,m$ are natural numbers which $m$ is divisible by 3 and $\frac{m}{3}$ is perfect square. $A, B$ and $C$ are matrices of size $n\times n$ and $P$ is a permutation of size $m$.
The inputs of $\ultang$ are $A$, $n_1$ and the output matrix $\ultang(A,n_1)$ is expressed as follows.\\

\begin{figure}
\centering
\includegraphics*[height=5cm, width=5.5cm]{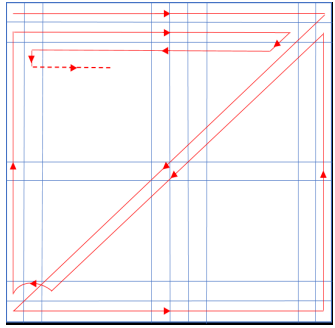}
\caption{
The path of one dimentional upper and lower triangular shift function ($\ultang$).
}
\label{ULTS}
\end{figure}

As seen in Figure \ref{ULTS}, the pixel shuffling by $\ultang$ starts from $A(1,1)$ entry and continues along the first row. Entries are shifted in the number of $n_1$ units. After reaching $A(1,n)$, it changes direction to the secondary diameter until it reaches $A(n,1)$. Upon reaching $A(n,1)$, entries keep on  moving across the last row and last column, respectively. In $A(2,n)$, the route again change to diagonal, and this time along the sub-secondary diagonal. In this shift, there are some mutations of different lengths over secondary diameters that the first and shortest jump is from $A(n-1,3)$ to $A(n-1,1)$. After first jump, we move straight from bottom to top towards $A(2,1)$ and after entering this entry, we move towards $A(2,n-3)$.  After a unit of diagonal movement, we enter the third row to traverse $A(3,2)$. Now we enter $A(4,2)$  and similarly continue the displacement of entries. Note that in this process, we do not enter a situation that we have already
passed through, and as mentioned before, we make jumps wherever necessary.\\

\begin{figure}
\centering
\includegraphics*[height=7.3cm, width=11cm]{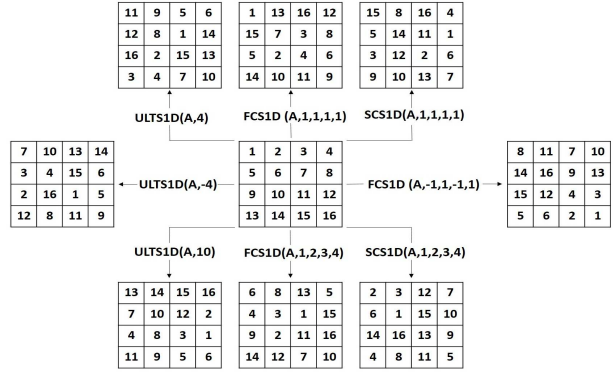}
\caption{
Examples for 1D shift functions.
}
\label{yyy}
\end{figure}

Now, by applying composition of functions and the shift function defined above, we will introduce $\fcs$ and $\scs$ functions. The outputs $B=\fcs(A,n_1,n_2,n_3,n_4)$ and $C=\scs(A,n_1,n_2,n_3,n_4)$ are obtained as follow.\\
\line(1,0){150}
\begin{algorithmic}
\State$B=\ultang(A,n_1)$;
\State$B=\ultang(B^\prime,n_2)$;
\State$B=\ultang\big(\fliplr(B),n_3\big)$
\State$B=\ultang\big(\flipud(B),n_4\big)$;

\State$C=\ultang\big(\flipud(A),n_1\big)$;
\State$C=\ultang\big(\fliplr(C),n_2\big)$;
\State$C=\ultang(C^\prime,n_3)$;
\State$C=\ultang(C,n_4)$;
\end{algorithmic}
\line(1,0){150}\\

\noindent which $B^\prime$ is transpose matrix of $B$. For better understanding,
examples for one-dimensional shift functions are given in Fig \ref{yyy}.

\begin{figure}
\centering
\includegraphics*[height=8cm, width=12cm]{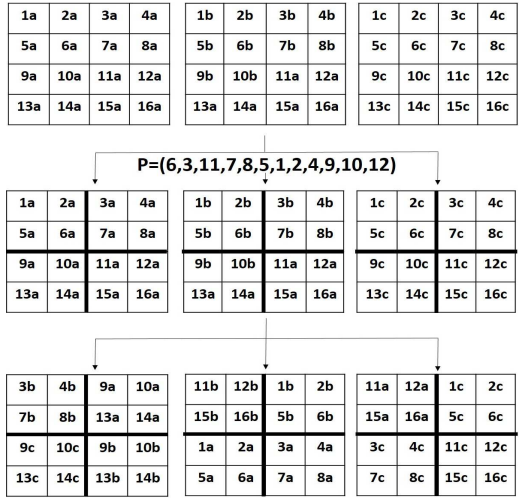}
\caption{
Example for the 3D shift function for $m=12$.
}
\label{yyy2}
\end{figure}

In order to shuffle pixels of matrices $A, B$ and $C$ by swapping their parts based on $P$, we define three-dimensional shift function which is shown by $\std$.
To get the output matrices of $\std(A,B,C,P)$, first, we block the matrices $A,B$ and $C$ to obtain the block matrices
$\tilde{A}, \tilde{B}, \tilde{C}\in(\mathbb{R}^{p\times p})^{\frac{n}{p}\times\frac{n}{p}}$, which $p=\frac{n}{\sqrt{\frac{m}{3}}}$.
Then, we convert these block matrices to block vectors $V\tilde{A}, V\tilde{B}, V\tilde{C}\in(\mathbb{R}^{p\times p})^{1\times\frac{n^2}{p^2}}$.
Next, we consider the block vector $\tilde{V}=[V\tilde{A}, V\tilde{B}, V\tilde{C}]\in(\mathbb{R}^{p\times p})^{1\times\frac{3n^2}{p^2}}$.
Now is the time to change the location of blocks of $\tilde{V}$ based on $P$ to get block vector $\tilde{W}\in(\mathbb{R}^{p\times p})^{1\times\frac{3n^2}{p^2}}$ as follow.\\
\line(1,0){150}
\begin{algorithmic}
\For {j=1:$\Sz$($\tilde{V}$,2)}
\State$\tilde{W}\{1,j\}=\tilde{V}\big\{1,P(1,j)\big\}$;
\EndFor
\end{algorithmic}
\line(1,0){150}\\
Output matrices of $\std(A,B,C,P)$ can be obtained by
dividing cell vector $\tilde{W}$ into three equal
 parts and deform them into matrices. In Fig. \ref{yyy2}, the steps of blocking and relocating the
 blocks under a specific permutation are given.

\begin{figure}
\centering
\includegraphics*[height=6cm, width=8cm]{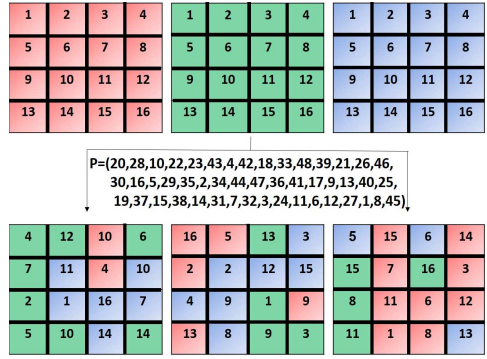}
\caption{
Example for the 3D shift function for $m=48$.
}
\label{yyy3}
\end{figure}

Also Fig \ref{yyy3} is the step of relocating blocks of block matrices for $m=48$.
All the above operations are invertible and inverses of $\fcs$ and $\scs$ and $\std$
are shown by $\invfcs$ and $\invscs$ and $\invstd$, respectively.


\section{Cellular Automata}\label{CAS}

\begin{figure}
\centering
\includegraphics*[height=7cm, width=12cm]{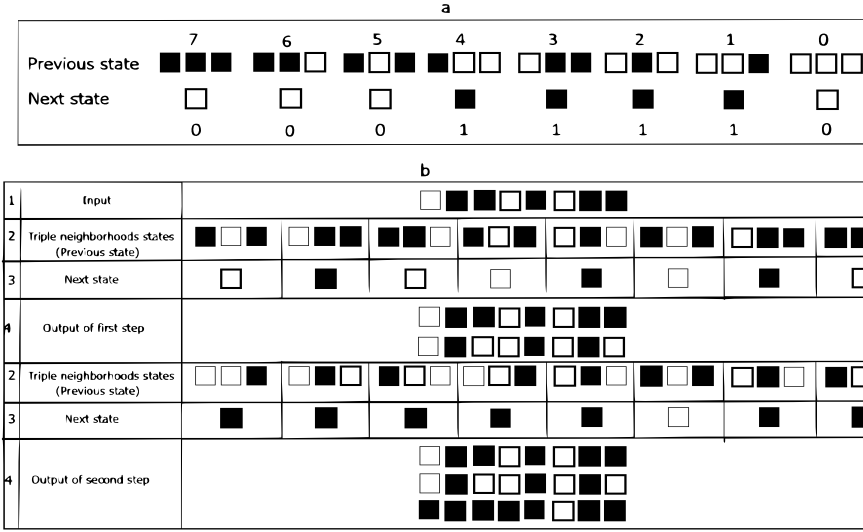}\\
\vspace{-0.6cm}
\emph{\caption{
(a) Cellular automata with rule number $30$, (b) Cellular automata process.
}\label{f1}}
\end{figure}

The study of cellular automata(CA) dates back to Von.Neumonn \cite{27}. CA is a model to describe a dynamic system composed of discrete cells where each cell can assume either the value $0$ or $1$.
These cells create a lattice which changes
in discrete time according local rules.
Generally, a CA can be defined as $\{C,S,V,F\}$, where $C$ is the cell space and $S$ is the discrete state sets, like $\{0,1\}$. $V$ determines cellular neighborhoods and $F$ is  transfer function\cite{28,29,30}. CA are categorized in different dimensions \cite{31} and the process of running one dimensional CA with local rule $30$ and triple neighborhoods are shown in Fig. \ref{f1}. At this fig,
1 and 0 are shown with black and white squares, respectively, and $[0101110]$ is considered as the first input of CA. Eight possible states for $[0101110]$ with triple neighborhoods are shown in the second row. Third row shows the next state of cells. This process up to three steps, is shown in the last row of Fig. \ref{f1}.
 Cellular automat
,which works based on simple logic computations, with pseudorandom hash behavior \cite{32},
is  highly parallel and distributed systems that can simulate complicated behaviors \cite{29}. One of the ways to create a strong cipher system is to use cellular automata with the large number of rules.\\

In our proposed algorithm, one-dimensional CA with
 triple neighborhoods is used like Fig. \ref{f1}, which the rule number for each input is obtained based on proposed chaotic system.

\begin{rem}
The outputs of the decryption by reversible cellular automata
is shown by
\begin{align*}
[\textbf{x}^\prime,\textbf{y}^\prime]:=\Phi( \textbf{x},  \textbf{y}, rule, rep),
\end{align*}
where  $\textbf{x}$
and $\textbf{y}$ are inputs in the $t-1$ and $t$ times, respectively.  "$rule$" is the rule number
after "$rep$" repetitions.\\
\end{rem}

In the proposed algorithm, the matrix $M_{n\times n}$ is encrypted based on reversible cellular automata by following algorithm. \\
\line(1,0){250}
\begin{algorithmic}
\State $R=\big[ {}_{2}^{\gamma}\Psi^{n/2}_r(1,:),
{}_{2}^{\gamma}\Psi^{n/2}_r(2,:),{}_{2}^{\gamma}\Psi^{n/2}_r(3,:),
 {}_{2}^{\gamma}\Psi^{n/2}_r(4,:)\big]$;
\For{$i=1:n$}
\If {$i==1$}
        \State $T=\diag(M,i)$;
        \State $X=T(1:n/2)$;
        \State $Y=T(n/2+1:n)$;
\ElsIf  {$i != 1$}
        \State $X=\diag(M,i)$;
        \State $Y=\diag(M,-i)$;
\EndIf
        \State $[\textbf{x},\textbf{y}]=\Phi( \textbf{x},  \textbf{y}, rule=R(i), 1);$
\EndFor
\end{algorithmic}
\line(1,0){250}\\

\section{Proposed encryption algorithm}
The details of the encryption algorithm are given in this section. The proposed
algorithm includes two parts. In the first part a sensitive algorithm for generating key space
is introduced and in the second part, an algorithm is proposed for image encryption.
\subsection{Key space generation algorithm}
In the process of encryption and decryption methods, the important
part that affects the output of the algorithm is the key space.
In this section, an algorithm to generate an efficient key space is introduced.
One of the most
important features of the proposed key space is the sensitivity to slight changes in
input images, which is evaluated in the section related to simulation results.
First, we design the proposed algorithm for image as follows. \\

\noindent\line(1,0){280}\\
\vspace{-0.3cm}
Algorithm 1: Key space generation algorithm for $I\in R^{n \times m}$.\\
\vspace{-0.3cm}
\noindent\line(1,0){280}\\
\begin{algorithmic}
\State  Input:  $I,r$;
\State  $r_0=4 \rand()$;
\State  $r_1:=\lfloor r_0 \times 10^3\rfloor$;
\State  $I_1=\circshift\big(\reshape(I^{\prime},1,[]),r_1\big)$;
\State  $\gamma_0:=I_1(1:4)$;
\State  $q:=^{\gamma_0}\Psi^1_r$;
\For{$i=7:3:nm$}
\State $\gamma_1:=\Big(\mean\big(q(2:4)\big),q(1),I_1(i-2),I_1(i-1),I_1(i)\Big)$;
\State $q=^{\gamma_1}\Psi^1_r$;
\EndFor
\State output: $r_0,q$;
\end{algorithmic}
\line(1,0){280}\\

This part of the algorithm is shown by the following notation according to the output and input values:
\begin{align*}
(r,r_0,q)=\key(I,r).
\end{align*}
For a color image $I_{n\times m \times 3}$, consider $I_1\in R^{n\times 3m}$ as follows.
\begin{align*}
I_1\big(1:n,(i-1)m+1:im\big):=I(:,:,i),~~i=1,2,3.
\end{align*}
Then define
\begin{align*}
I_2=\circshift\big( I_1,\lfloor r_0 \times 10^3\rfloor \big),
\end{align*}

where $r_0:=4 \rand()$.  By using Algorithm 1, obtain

\begin{align*}
(r,r_i,q_i)=\key\Big(I_2\big(1:n,(i-1)m+1:im\big),r\Big),~~i=1,2,3.
\end{align*}

In the last step of key generation for color image, consider key space as
\begin{align*}
(r_0,q)=\Big(\mean\big(\{r_i\}^{3}_{i=1}\big),\mean\big(\{q_i\}^{3}_{i=1}\big) \Big).
\end{align*}
\begin{figure}[h!]
\centering
\subfigure{
\includegraphics*[width=0.75\textwidth]{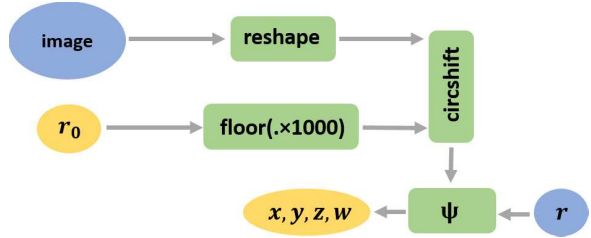}}\\
\vspace{-1.0cm}
\emph{\caption{
The structure of the proposed key generation algorithm.}\label{f9}}
\end{figure}
\subsection{Encryption and decryption algorithm}
The steps of the proposed encryption algorithm for a grayscale image $I\in R^{n \times m}$ based on the new proposed chaos system are written as follows.\\

\textbf{step 1.} By using key space generation algorithm obtain:
\begin{align*}
(r,r_0,q)=\key(I,r),
\end{align*}
where $q=(x,y,z,w)$.\\

\textbf{step 2.} Define $\textbf{v}_1 \in R^{1 \times n}$, $\textbf{v}_2 \in R^{1 \times 8m}$
and $M \in R^{n \times m}$as
\begin{align*}
&\textbf{v}_1:=\lfloor \chi^n_r \times 10^{3}\rfloor,\\
&\textbf{v}_2:=\lfloor \chi^{8m}_{r_0}\times 10^{3}\rfloor,\\
&M:=mod\Big( \big \lfloor \reshape \big( \chi^{nm}_{\mean([r,r_0])} \times 10^{3},[n,m] \big)\big \rfloor,255\Big),\\
\end{align*}
where
\begin{align*}
\chi^j_r=\big[^{q}\Psi^{\frac{j}{4}}_r(1,:),~~^{q}\Psi^{\frac{j}{4}}_r(2,:),~~^{q}\Psi^{\frac{j}{4}}_r(3,:),~~^{q}\Psi^{\frac{j}{4}}_r(4,:)\big],
~~j=n,8m,nm.
\end{align*}

\textbf{step 3.}  Transfer $I$  to binary image $I_1 \in R^{n \times 8m}$, and shift as

\line(1,0){250}
\begin{algorithmic}
\For {i=1:n}
\State $I_1(i,1:8m)=\circshift\Big(\big(I_1(i,1:8m)\big)',\textbf{v}_1(i)\Big)'$;
\EndFor
\For {i=1:4}
\State $R\big((i-1)n+1:2n,1:2m\big)=I_1\big(1:n,2(i-1)m+1:2im\big)$;
\EndFor
\For {i=1:2m}
\State $SR(1:4n,i)=\circshift\big(R(1:4n,i),\textbf{v}_2(i)\big)$;
\EndFor
\For {i=1:4}
\State $I_2\big(1:n,2(i-1)m+1:2im\big)=SR\big((i-1)n+1:in,1:2m\big)$;
\EndFor
\end{algorithmic}
\line(1,0){250}\\

\textbf{step 4.} Consider most and least significant bits in two separate matrices $I_2^{m},I_2^{l}\in R^{n\times 4m}$, respectively. Then define $I_3 \in R^{s \times 4m}$ with $s=n/2-mod(n/2,3)$ by
\begin{align*}
I_3=I^{m}_{2}(1:2:2s,:).
\end{align*}

\textbf{step 5.} By using proposed algorithm based on cellular automata in Section \ref{CAS}, obtain
$I_4$ for input values $I_3,q$ and $r$.

\textbf{step 6.} By inverting steps 3 and 4 on $I_4$
obtain  $I_5 \in R^{n \times m}$ and shift $I_5$ as

\begin{align*}
I_6=\fcs \big(I_5,\textbf{v}_1(10),-\textbf{v}_1(20),\textbf{v}_1(30),-\textbf{v}_1(40)\big).
\end{align*}

\textbf{step 7.} Obtain final encryption image $EI$ by using bitwise XOR operation as
\begin{align*}
EI=I_6\oplus M\oplus M_s,
\end{align*}
where $M_s:=\circshift \Big(M,\big[\lfloor q(1) \rfloor, \lfloor q(0)\rfloor \big]\times10^2\Big)$.\\

Now, we develop the above proposed algorithm for color image $I\in R^{n \times m \times 3}$. In the first step, by using key space algorithm for color image obtain $(r,r_0,q)$. Next
by  $\std$ shift function obtain $I_1$ as
\begin{align*}
 \big( I_1(:,:,1),I_1(:,:,2),I_1(:,:,3)\big) =\std\big( I(:,:,1),I(:,:,2),I(:,:,3),p\big),
\end{align*}
where $p=\sort(^{q}\Psi^{nm}_r)$.
Finally, each layer of $I_1$ is encrypted using the encryption algorithm that described above.\\
Decryption processes are achieved simply by following reverse steps of the above algorithms.

\begin{figure}[h!]
\centering
\subfigure{
\includegraphics*[width=0.9\textwidth]{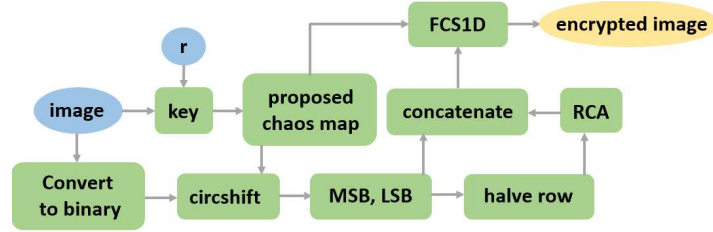}}\\
\vspace{-0.9cm}
\emph{\caption{The structure of the proposed encryption algorithm.}\label{Fs18}}
\end{figure}

\section{Simulation results}

In this section, we review the results of the algorithm and
to check the security of the proposed algorithm,
we perform different types of security tests.
\subsection{Key space analysis}
In order to show the sensitivity of the key generation algorithm to the input values,
Table \ref{tkey} shows the output results of the algorithm for different inputs.
According to the results, it can be seen that small changes in the input of the algorithm have caused changes in the key space. To withstand the brute force attacks
the key space must be large enough. According to \cite{kke}, this size must be greater than
$2^{100}$. If the precision computing is considered as $10^{-15}$, then the key space
size is $10^{90}=2^{90log_{2}10} \approx 2^{298}$, and this is large enough to resist
the attack.

\begin{center}
\begin{table}[ht!]
\footnotesize
\caption{ {\footnotesize Key space generated by the proposed algorithm for various inputs and $r=0.7$.}}
\label{tkey}
\centering
\begin{tabular}{cccccccccccc}
\hline
          \multicolumn{2}{c}{}   &\multicolumn{5}{c}{Keys }   \\\cmidrule{3-7}
Image&   &$r_0$                       &$x$     &$y$   &$z$    &$w$     \\
\hline \hline
lena     &  Original        &\verb"constant(0.7)"       &0.6038  &0.2828    &0.7086   &0.4590     \\
$(256\times256\times3)$  &  One bit changed &\verb"constant(0.7)"         &0.5495  &0.3590    &0.9294   &0.3930     \\
           &   Original        &\verb"random"                  &0.6600  &0.7191    &0.3064   &0.8206      \\
           &  One bit changed &\verb"random"                    &0.4336  &0.4298    &0.2942   &0.4056     \\
           &  First time run        &\verb"random"                &0.2499  &0.4357    &0.5456   &0.5322       \\
           &  Second time run   &\verb"random"                   &0.4351  &0.2396    &0.6711   &0.5193    \\
           \hline
lena    &  Original        &\verb"constant(0.7)"                 &0.7753  &0.3087    &0.5087   &0.1720       \\
$(256\times256)$   &  One bit changed &\verb"constant(0.7)"       &0.6800  &0.6547    &0.2556   &0.3093    \\
           & Original        &\verb"random"                                      &0.8145  &0.4966    &0.3931   &0.0714     \\
           &  One bit changed &\verb"random"                              &0.0451  &0.4445    &0.0295   &0.7943     \\
           \hline
pirate         & Original        &\verb"constant(0.7)"                                &0.3329  &0.9476    &0.2881  &0.5595       \\
$(256\times256)$         &   First pixel changed     &\verb"constant(0.7)"  &0.8755  &0.1108    &0.6811  &0.4796      \\
                                  &   Last pixel changed     &\verb"constant(0.7)"  &0.3318  &0.6714    &0.1125  &0.6366      \\
           &  Original        &\verb"random"                                              &0.8654  &0.2863    &0.0157  &0.5612       \\
           &  First pixel changed        &\verb"random"                               &0.7584  &0.9722    &0.1689  &0.9224       \\
          &  Last pixel changed        &\verb"random"                               &0.7952  &0.4099     &0.1525  &0.0080       \\
           \hline
\end{tabular}
\end{table}
\end{center}

\subsection{Statistical analysis}
\begin{figure}[h!]
\centering
\subfigure{
\includegraphics*[width=1\textwidth]{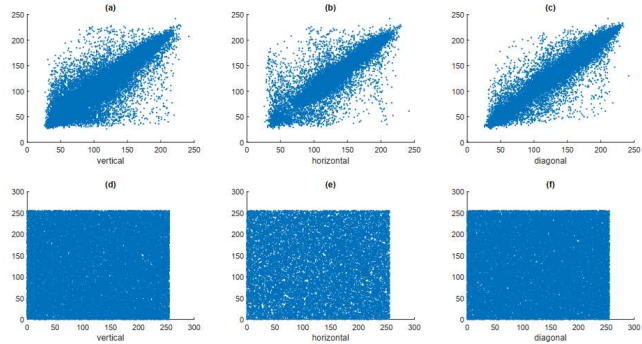}}\\
\vspace{-0.5cm}
\emph{\caption{
Correlation of neighborhood pixels at different directions before (a-c) and after encryption
(d-f) of lena ($256 \times 256$). }\label{Fs1}}
\end{figure}

\begin{figure}[h!]
\centering
\subfigure{
\includegraphics*[width=1.2\textwidth]{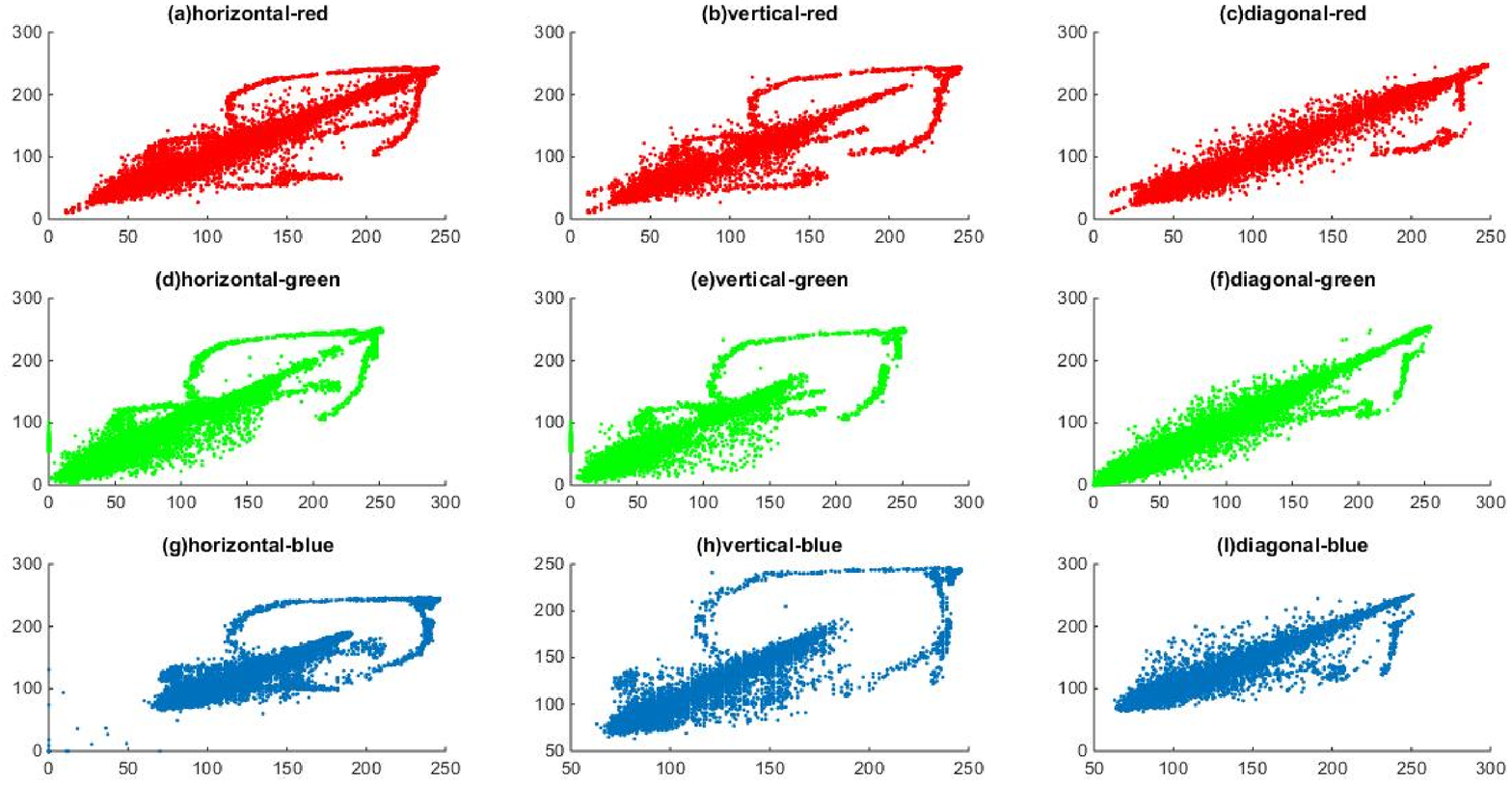}}\\
\vspace{-0.9cm}
\emph{\caption{
Correlation of neighborhood pixels at different directions for  4.1.04. }\label{Fs2}}
\end{figure}

\begin{figure}[h!]
\centering
\subfigure{
\includegraphics*[width=1.2\textwidth]{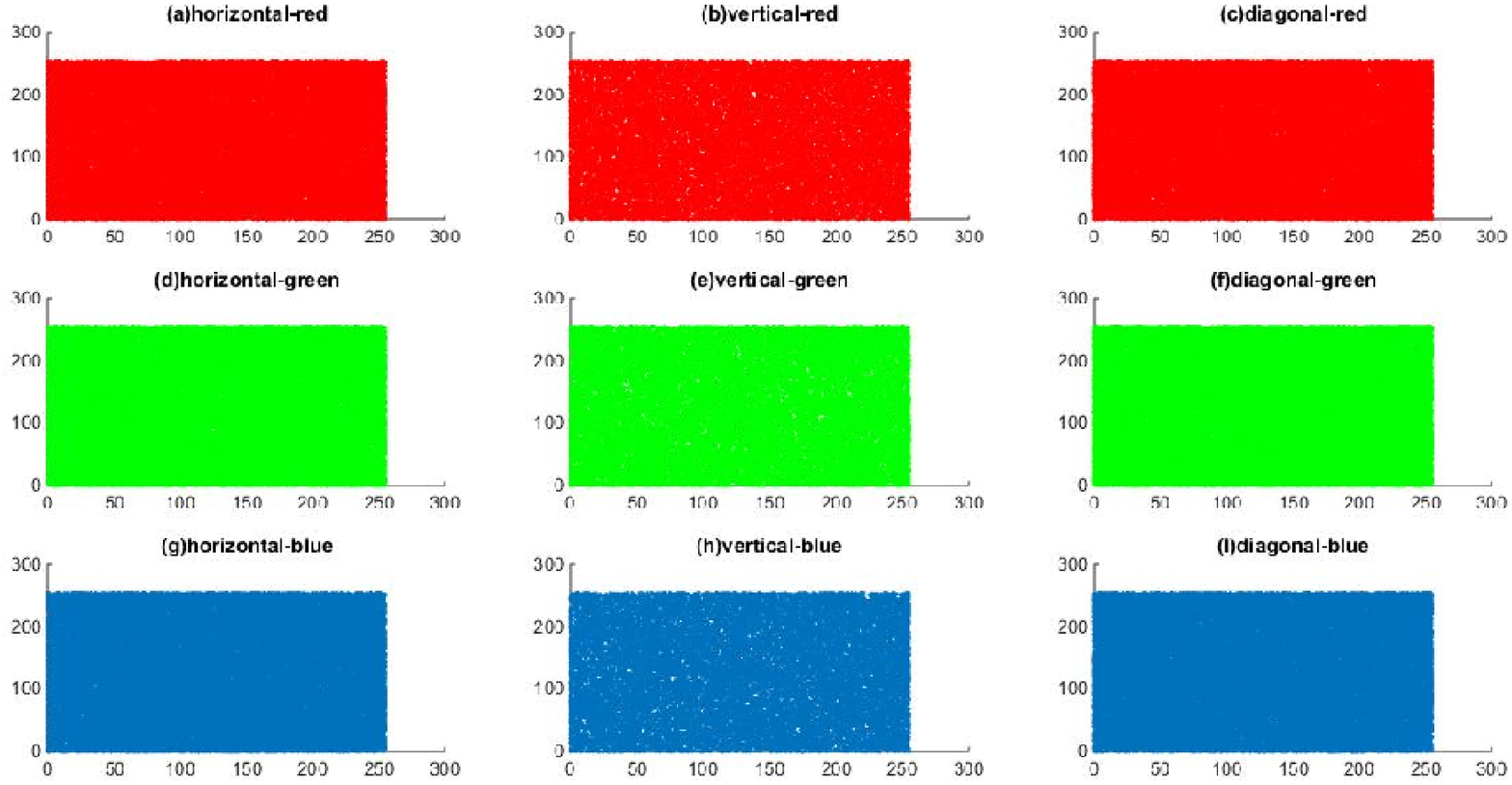}}\\
\vspace{-0.9cm}
\emph{\caption{
Correlation of neighborhood pixels at different directions for encrypted 4.1.04 }\label{Fs3}}
\end{figure}

Statistical test consists of three basic tests,
 correlation values, information entropy and histogram analysis.
The correlation values for an image are calculated using the following formula
\begin{align*}
C_{x,y}=\frac{E(x-\bar{x})(y-\bar{y})}{\sigma_x \sigma_y},
\end{align*}
where $E,\bar{x},\bar{y},\sigma_x$ and $\sigma_y$ represent
expectation, mean values and standard deviation, respectively.
If this number is close to zero, the encrypted image will not have significant information from the original image. The results of this value for original and encrypted image are give in
Tables \ref{T1} and \ref{T2}. By comparing values of encrypted and original images, it is observed that the results for the encrypted images are close to the ideal value, i.e., zero.
Also the correlation distributions
for the original and encrypted images are shown in Figures \ref{Fs1}-\ref{Fs3}.
By using these figures, it can be seen that encrypted images have
more uniform distribution.

\begin{center}
\begin{table}[ht!]
\footnotesize
\caption{{\footnotesize Correlation coefficient and Information entropy values in the plain and cipher grayscale images.}}
\label{T1}
\centering
\begin{tabular}{c@{\hspace*{0.1cm}}c@{\hspace*{0.1cm}}c@{\hspace*{0.1cm}}c@{\hspace*{0.1cm}}c@{\hspace*{0.1cm}}c@{\hspace*{0.1cm}}
c@{\hspace*{0.1cm}}c}
\hline
           &\multicolumn{4}{c}{Correlation coefficient}&  Information \\\cmidrule(l){2-5}
           &Horizontal  &Vertical&Diagonal&Diagonal& entropy\\
           & &&\scriptsize{(lower left to top right)}& \scriptsize{(lower right to top left)}&\\

\hline
\textbf{plain images} && &  &&&&\\
5.1.12                 &0.95649 &0.97408 &0.93627&0.93893&6.7057\\
5.2.08                 &0.93707 &0.89264 &0.85427&0.85572&7.201\\
boat($512\times 512$)  &0.93812 &0.97131 &0.92585&0.92216&7.1914\\
lena($256\times 256$)  &0.92578 &0.95926 &0.92577&0.90374&7.4429 \\
lena($512\times 512$)  &0.20131 &0.21056 &0.19815&0.19016&7.4455\\
\hline
\textbf{encrypted images }&& &  &&&\\
5.1.12                 &$-0.00047766$&$0.00061399$&$0.001538$&$-0.0001677$&7.9973\\
5.2.08                 &$-0.0003788$ &0.00013229 &$-0.00082042$&$0.00027628$&7.9993\\
boat($512\times 512$)  &0.00093687   &0.00057866 &0.00038615   &$-0.00084275$&7.9993\\
lena($256\times 256$)  &$-0.0021188$ &$-0.0025797$ &$-0.0015225$&0.00026687&7.9972\\
lena($512\times 512$)  &0.00039856  &0.00015902 &$-0.00023439$&$0.00069882$&7.9993\\
\hline
\end{tabular}
\end{table}
\end{center}
\begin{center}
\begin{table}[ht!]
\footnotesize
\caption{{\footnotesize Correlation coefficient values in the plain and cipher grayscale images.}}
\label{T2}
\centering
\begin{tabular}{c@{\hspace*{0.08cm}}c@{\hspace*{0.1cm}}c@{\hspace*{0.1cm}}
c@{\hspace*{0.1cm}}c@{\hspace*{0.1cm}}c@{\hspace*{0.1cm}}cccccc}
\hline
            &&\multicolumn{4}{c}{Correlation coefficient}&  Information \\\cmidrule(l){3-6}
           &&Horizontal  &Vertical&Diagonal&Diagonal&entropy\\
           && &&\scriptsize{(lower left to top right)}& \scriptsize{(lower right to top left)}&\\

\hline
\textbf{plain images} & && &  &&&&\\
                       &R&0.94928 &0.95616 &0.91631&0.91764&6.2499\\
4.1.02                 &G&0.93077 &0.95338 &0.8968&0.90017&5.9642\\
                       &B&0.91784 &0.94421 &0.88698&0.88898&5.9309 \vspace{0.1cm} \\
                       &R&0.97865 &0.9879 &0.96875&0.96841&7.2549\\
4.1.04                 &G&0.96598 &0.98201 &0.95147&0.95072&7.2704\\
                       &B&0.95231 &0.97178 &0.93069&0.93057&6.7825 \vspace{0.1cm} \\
                       &R&0.92307 &0.86596 &0.85187&0.85434&7.7067\\
4.2.03                 &G&0.86548 &0.76501 &0.72493&0.7348&7.4744\\
                       &B&0.90734 &0.88089 &0.84244&0.83986&7.7522 \vspace{0.1cm} \\
                       &R& 0.97527    &0.98531 &0.97339 &0.96484&5.0465\\
lena                   &G&0.96662 &0.98017 & 0.96303 & 0.95357&5.4576\\
($512\times512\times3$)&B& 0.93339    &0.95579 & 0.92643&0.91863&4.8001\\
\hline
\textbf{encrypted images }&&& &  &&&&\\
                       &R&0.0011619  &0.0016396     &0.00053731&0.00082608&7.9973\\
4.1.02                 &G&0.0007757  &$-0.00021986$ &0.00069429&0.00088685&7.9971\\
                       &B&0.00029618 &0.0005109     &0.00098397&$-0.0015792$&7.9972 \vspace{0.1cm} \\
                       &R& 0.00036129 &$-0.00047313$ &-0.0015582&0.0026443&7.9973 \\
4.1.04                 &G&0.0029232 &0.0014998 &0.0010652&0.001758&7.9972 \\
                       &B&0.0031419 & 0.00095978 &0.0019824&$-0.001847$&7.9973 \vspace{0.1cm} \\
                       &R&$-0.00057646$ & 0.00020068 & 0.0010341&$-0.00034744 $&7.9993\\
4.2.03                 &G&0.00053615 &$-0.00029491$&0.00061376 &$-0.00040553$&7.9993\\
                       &B&$-0.00080725$ &$-0.0017738$&$-0.0002713$&0.0009324&7.9993 \vspace{0.1cm} \\
                       &R&$-2.2351e-06$ & 0.00022357 & $-0.00074232$&$-0.00055947 $&7.9993\\
lena                   &G&0.0002728 &$-3.7743e-05 $& 0.0012758&$-0.00036377$&7.9993\\
($512\times512\times3$)&B&0.00031603 &$0.00067082$&$-0.00071107$&0.0012129&7.9993\\

\hline
\end{tabular}
\end{table}
\end{center}

\begin{figure}[h!]
\centering
\subfigure{
\includegraphics*[width=1.0\textwidth]{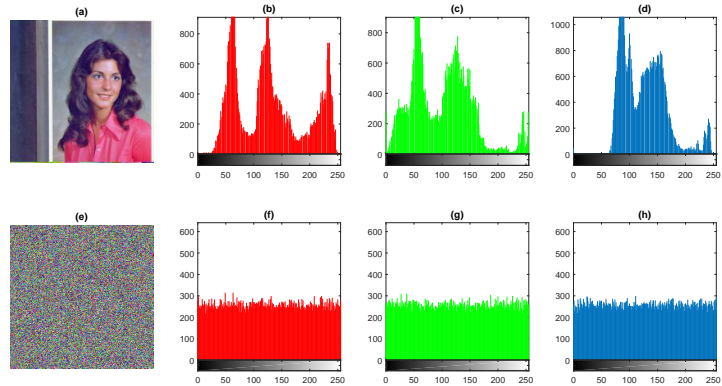}}\\
\vspace{-0.9cm}
\emph{\caption{ (a)-(d) the original image and its histogram, (e)-(f) the output image of the proposed algorithm and its histogram}\label{Fs4}}
\end{figure}

\begin{figure}[h!]
\centering
\subfigure{
\includegraphics*[width=1.0\textwidth]{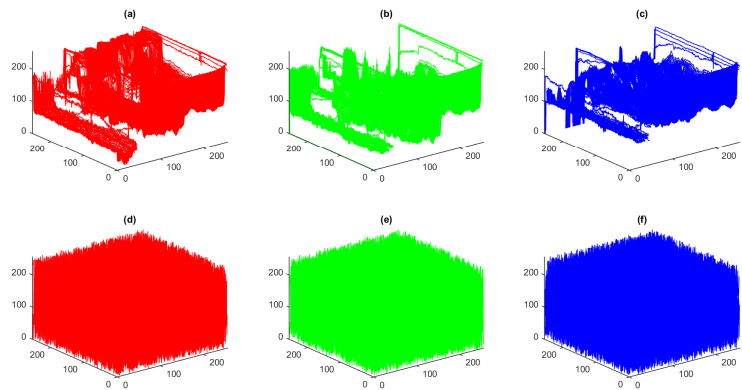}}\\
\vspace{-0.9cm}
\emph{\caption{(a)-(c) 3D histogram representation of original image and (d)-(f)  encrypted image for
4.1.04. }\label{Fs5}}
\end{figure}

The uncertainty of information for an image is evaluated
by using Shannon information entropy. This value is calculated by
\begin{align*}
H=-\sum_{i} p(i)\log\big(p(i)\big),
\end{align*}
where $p(i)$ is the probability of occurrence of each pixel.
 The ideal value for this test is 8.
If the results of the algorithm are close to this number,
it will be difficult to obtain valid information
from the image. The results of this test for different images with different sizes are given
in Tables \ref{T1} and \ref{T2}. The results in these tables show that the information entropy
of encrypted image by proposed algorithm are close to the ideal value.
The distribution characteristics of pixel values is evaluated by histogram.
Histograms of plaintext and encrypted images are given in Figure \ref{Fs4}.
Also the intensity histogram is shown in Figure \ref{Fs5}. Using these results, it can be seen that
the histograms of the cipher image are uniformly distributed in comparison
 with the original
image. Therefore, according to the results the proposed algorithm has a good ability to resist statistical attacks.
\subsection{Differential attack test}
One of the most important attacks evaluated on cryptographic algorithms is differential attacks.
This type of attack is studied by two tests,
UACR test (number of pixels change rate) and NPCI test (unified average changing
intensity). These values are calculated  by

\begin{align*}
&NPCR=\frac{\sum_{i,j}D_{i,j}}{n\times m} \times 100\%, \\
&UACI=\frac{\sum_{i,j}|C^1_{i,j}-C^2_{i,j}|}{n \times m}\times 100\%,
\end{align*}
where $C^1$ and $C^2$ are two encrypted images such that their plain images differ
only by one pixel. Also if $C^1_{i,j}\neq C^2_{i,j}$ then  $D_{i,j}=1$ otherwise $D_{i,j}=0$.
The ideal values for these tests are  $100\%$ and $33.\overline{33}\%$, respectively.
The results are shown in Tables \ref{T3}-\ref{T5}.
The best study on these values is discussed in \cite{7}, which critical intervals for NPCR and UACI are considered. In Tables \ref{T3} and \ref{T4}, the results of the proposed algorithm for different images are compared to the critical values. These results indicate the success of the proposed method in these tests.
Also in Table \ref{T5}, the proposed method are compared to other studies.

\begin{center}
\begin{table}[ht!]
\footnotesize
\caption{{\footnotesize Numerical results of the UACI and NPCR for different images ($256 \times 256$).}}
\label{T3}
\centering
\begin{tabular}{c@{\hspace*{0.1cm}}c@{\hspace*{0.1cm}}c@{\hspace*{0.1cm}}c@{\hspace*{0.1cm}}c@{\hspace*{0.1cm}}c
@{\hspace*{0.1cm}}c@{\hspace*{0.1cm}}c@{\hspace*{0.1cm}}c@{\hspace*{0.1cm}}c@{\hspace*{0.1cm}}c@{\hspace*{0.1cm}}cc}
\hline
\multicolumn{3}{c}{}   &\multicolumn{3}{c}{UACI critical values \cite{7} }& &\multicolumn{3}{c}{NPCR critical values \cite{7} }\\ \cmidrule{4-6}\cmidrule{8-10}
                     &&&u$^{*-}_{0.05}$=33.2824&u$^{*-}_{0.01}$=33.2255& u$^{*-}_{0.001}$=33.1594
&&N$^{*}_{0.05}=$&N$^{*}_{0.01}=$&N$^{*+}_{0.001}=$
                       \\
Image                 &&UACI&u$^{*+}_{0.05}$=33.6447&u$^{*+}_{0.01}$=33.7016&u$^{*+}_{0.001}$=33.7677
&NPCR&99.5693&99.5527&99.5341 \\
\hline
5.1.12                &&33.4570&Pass&Pass&Pass&99.6127&Pass&Pass&Pass    \\
lena                  &&33.4931&Pass&Pass&Pass&99.6112&Pass&Pass&Pass  \vspace{0.1cm}    \\
                      &R&33.4931 &Pass&Pass&Pass&99.6213 &Pass&Pass&Pass     \\
4.1.02                &G&33.4763&Pass&Pass&Pass&99.6000&Pass&Pass&Pass    \\
                      &B&33.4421&Pass&Pass&Pass&99.6147&Pass&Pass&Pass \vspace{0.1cm} \\
                      &R&33.4882 &Pass&Pass&Pass&99.6191     &Pass&Pass&Pass     \\
4.1.04                &G&33.4581&Pass&Pass&Pass&99.6032&Pass&Pass&Pass    \\
                      &B&33.5000&Pass&Pass&Pass&99.6001&Pass&Pass&Pass\\
\hline
\end{tabular}
\end{table}
\end{center}


\begin{center}
\begin{table}[ht!]
\footnotesize
\caption{{\footnotesize Numerical results of the UACI and NPCR for different images ($512 \times 512$).}}
\label{T4}
\centering
\begin{tabular}{c@{\hspace*{0.1cm}}c@{\hspace*{0.1cm}}c@{\hspace*{0.1cm}}c@{\hspace*{0.1cm}}c@{\hspace*{0.1cm}}c
@{\hspace*{0.1cm}}c@{\hspace*{0.1cm}}c@{\hspace*{0.1cm}}c@{\hspace*{0.1cm}}c@{\hspace*{0.1cm}}c@{\hspace*{0.1cm}}}
\hline
\multicolumn{3}{c}{}   &\multicolumn{3}{c}{UACI critical values \cite{7} }&&
\multicolumn{3}{c}{NPCR critical values\cite{7} }
\\\cmidrule{4-6}\cmidrule{8-10}
                       & &&u$^{*-}_{0.05}$=33.3730&u$^{*-}_{0.01}$=33.3445& u$^{*-}_{0.001}$=33.3115
&&N$^{*}_{0.05}=$&N$^{*}_{0.01}=$&N$^{*+}_{0.001}=$
                         \\
Image                &&UACI&u$^{*+}_{0.05}$=33.5541&u$^{*+}_{0.01}$=33.5826&u$^{*+}_{0.001}$=33.6156&
NPCR &99.5893&99.5810&99.5717
\\
\hline
5.2.08                &&33.4571&Pass&Pass&Pass&99.6115&Pass&Pass&Pass    \\
boat                  &&33.4750&Pass&Pass&Pass&99.6137&Pass&Pass&Pass    \\
lena                  &&33.4683&Pass&Pass&Pass&99.6046&Pass&Pass&Pass  \vspace{0.1cm}   \\
                      &R&33.4546 &Pass&Pass&Pass&99.6215 &Pass&Pass&Pass    \\
4.2.03               &G&33.4681&Pass&Pass&Pass&99.6064&Pass&Pass&Pass   \\
                      &B&33.4502&Pass&Pass&Pass&99.6110&Pass&Pass&Pass\vspace{0.1cm} \\
                      &R&33.4641 &Pass&Pass&Pass&99.6112 &Pass&Pass&Pass    \\
lena                 &G&33.4541&Pass&Pass&Pass&99.6080&Pass&Pass&Pass    \\
                      &B&33.4543&Pass&Pass&Pass&99.6101&Pass&Pass&Pass\\
\hline
\end{tabular}
\end{table}
\end{center}
\begin{center}
\begin{table}[ht!]
\footnotesize
\caption{{\footnotesize Correlation coefficient and Information entropy values in the plain and cipher grayscale images.}}
\label{T5}
\centering
\begin{tabular}{c@{\hspace*{0.3cm}}c@{\hspace*{0.3cm}}c@{\hspace*{0.3cm}}c@{\hspace*{0.3cm}}c@{\hspace*{0.3cm}}cc
@{\hspace*{0.3cm}}c@{\hspace*{0.3cm}}c@{\hspace*{0.3cm}}c@{\hspace*{0.3cm}}c@{\hspace*{0.3cm}}c}
\hline
           &\multicolumn{4}{c}{NPCR}&\multicolumn{4}{c}{UACI} \\\cmidrule(l){2-5}\cmidrule(l){6-10}
image &Propose  &Ref.\cite{1}&Ref.\cite{2}&Ref.\cite{3}&&Propose  &Ref.\cite{1}&Ref.\cite{2}&Ref.\cite{3}\\
\hline
5.1.09    &99.6117   &99.6078&99.6016&99.6064& &33.4490 &33.4563&33.4700&33.4456\\
5.1.10    &99.6164   &99.6098&99.6191&99.6154& &33.4733 &33.4510&33.4826&33.4946\\
5.1.11    &99.6122   &99.6077&99.6042&99.6244& &33.4654 &33.4832&33.5648&33.5541\\
5.1.14    &99.6118   &99.6129&99.6199&99.6364& &33.5111 &33.4848&33.4725&33.4655\\
7.1.01    &99.6161   &99.6040&99.6053&99.5992& &33.4682 &33.4779&33.4820&33.5037\\
7.1.02    &99.6093   &99.6016&99.6080&99.6075& &33.4730 &33.4172&33.4357&33.4237\\
7.1.09    &99.6152   &99.6061&99.6112&99.6162& &33.4591 &33.4814&33.4596&33.4177\\
7.1.10    &99.6110   &99.6052&99.6106&99.6045& &33.4817 &33.4852&33.4538&33.4344\\
\hline
\end{tabular}
\end{table}
\end{center}
\subsection{Noise and data loss attack}

\begin{figure}[h!]
\centering
\subfigure{
\includegraphics*[width=0.8\textwidth]{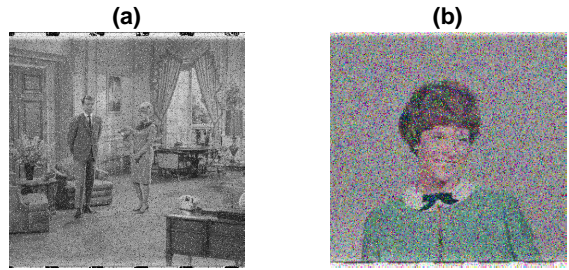}}\\
\vspace{-0.9cm}
\emph{\caption{Salt $\&$ pepper noise attack results with different noise density and extracted secret images:
(a), (d) decrypted images with noise intensity 0.1 and 0.2, respectively.}\label{Fs16}}
\end{figure}

\begin{figure}[h!]
\centering
\subfigure{
\includegraphics*[width=1.0\textwidth]{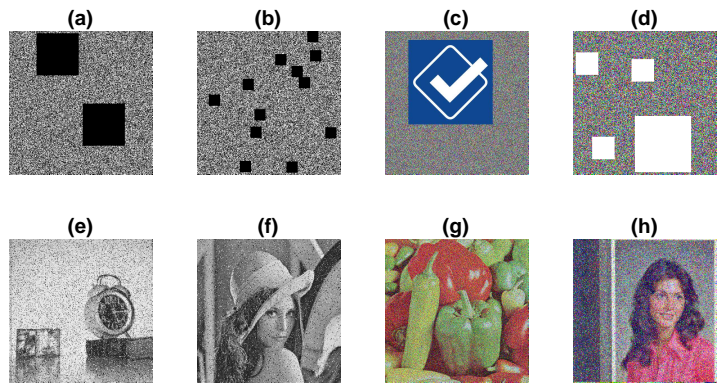}}\\
\vspace{-0.9cm}
\emph{\caption{Results of extracting the secret image after slight change in the key:
(a) almost 0.17 data loss, (b) almost 0.07 data loss, (c) almost 0.35 data loss, (d) almost 0.23 data loss,
(e)-(h) decrypted images.}\label{Fs17}}
\end{figure}

In the transmission of images over the network and through physical channels, part of the data is naturally or intentionally lost due to noise and cropped attacks. Therefore, efficient cryptographic schemes are capable of retrieving the original images even after these types of attacks.
In order to test the anti-noise ability of the proposed method, the decrypted images after the noise attack with different intensities $0.1$ for gray and $0.2$ for color images are given in Fig.\ref{Fs16}. The strength of the proposed method against cropped attacks with varying degrees of data loss is
 shown in Fig.\ref{Fs17}.

\section{Conclusions}
Create a secure environment for transferring images a burgeoning
subject branch. To study this topic, in the first step,
 by the development and improvement of the chaos system, a new chaos system is introduced.
 The structure of the new system is studied with different tests and the results show the efficiency of the new
 system. In the next step, the proposed system is used to create a secure
 image transfer environment in the form of an encryption algorithm.
 The results of studying the proposed algorithm by various security tests
 show that the proposed algorithm is efficient and safe.\\

\noindent\textbf{Funding} This work does not receive any funding.
\section*{Compliance with ethical standards}
\noindent \textbf{Conflict of interests} The authors declare that they have no conflict
of interests.\\
\noindent\textbf{Human participants and animals} This paper does not include
human participants and animals.


\end{document}